\documentclass[11pt,a4paper]{article}

\usepackage[T1]{fontenc}
\usepackage[utf8]{inputenc}
\usepackage[sc]{mathpazo}   % Palatino text and math
\usepackage{textcomp}
\usepackage{titling}

\usepackage{graphicx}
\usepackage{amsmath}
\usepackage{xurl}
\usepackage%[hidelinks]
{hyperref}
\hypersetup{								% hyperlinked styles
	colorlinks=true,      		            % false: boxed links; true: 
	linkcolor=blue,                        	% defined separately
	filecolor=blue,                     	% defined separately
	citecolor=blue,    	        			% defined separately
	urlcolor=blue,                 	   		% color of external links
	pdffitwindow=false,               	 	% window fit to page when opened
	pdfstartview={FitH},               		% fits the width of the page
	pdfauthor={Huw Price},           		% Author
}
\usepackage{tikz}
\usetikzlibrary{shapes,decorations,arrows,calc,arrows.meta,fit,positioning}
\tikzset{
    -Latex,auto,node distance =1 cm and 1 cm,semithick,
    state/.style ={ellipse, draw, minimum width = 0.7 cm},
    point/.style = {circle, draw, inner sep=0.04cm,fill,node contents={}},
    bidirected/.style={Latex-Latex,dashed},
    el/.style = {inner sep=2pt, align=left, sloped}
}

\usepackage{rotating,stackengine,scalerel}
\newcommand\wye{\scalerel*{\stackengine{-1pt}{%
  \rotatebox[origin=c]{30}{\rule{10pt}{.9pt}}\kern-1pt%
  \rotatebox[origin=c]{-30}{\rule{10pt}{1.3pt}}}{%
  \rule{.9pt}{10pt}}{O}{c}{F}{F}{S}}{\Delta}}

\begin{document}

\title{Bell Correlations and Selection Bias}
\author{Huw Price\thanks{Trinity College, Cambridge, UK; email \href{mailto:hp331@cam.ac.uk}{hp331@cam.ac.uk}.}}
\date{\today}
%\maketitle\thispagestyle{empty}%remove page number from title page

\begin{titlingpage}
    \maketitle
   \begin{abstract}
\noindent  
Methods of selecting samples from larger populations may produce bias, in either direction: inducing correlations between variables independent in the full population, or masking correlations between variables dependent in the full population. Here we propose a surprising application of these familiar ideas. We argue that they are relevant to puzzling correlations uncovered in quantum theory by John Stewart Bell (Bell 1964). In the light of Bell’s work and subsequent experiments it is widely believed that the quantum world is ‘nonlocal’, in apparent tension with relativity. Many hold that the only alternative is to abandon ‘realism’, the view that there is an objective world independent of measurement. We propose instead that Bell’s correlations are selection artefacts, in tension neither with relativity nor realism. In standard two-particle Bell experiments the relevant selection is a preselection, achieved by the preparation of the initial state. Again, selection bias via preselection is familiar elsewhere in science, but it doesn't seem to have been noticed that it is applicable in this case.\\

\end{abstract}
\end{titlingpage}

%\maketitle

\section{Introduction}

Selection artefacts are common in science. Procedures for selecting samples from larger populations may produce bias, in either direction: inducing correlations between independent variables, or masking correlations between dependent variables. Missing such bias has practical consequences, where statistics guide policy.

The case discussed here is a long way from policy, but of unusual theoretical interest. Some of the most puzzling correlations in modern science are those uncovered by the physicist John Stewart Bell in the 1960s (Bell 1964). Bell argued that if the predictions of quantum mechanics (QM) are correct, then there are correlations in the quantum world that cannot be explained by a \emph{local} theory -- that is, intuitively, by a theory in which causal influences propagate at finite speed, as required by special relativity. The relevant predictions have since been confirmed in many experiments.

Do these experiments prove that the world is \emph{nonlocal} in Bell's sense? It is often said that the only alternative is to reject \emph{realism} -- roughly, the view that the world exists independently of our measurement choices. The experiments in question are described as disproving \emph{local realism} -- as showing that either locality or realism must be abandoned.\footnote{(Shalm et al 2015) claims `a loophole-free violation of local realism', for example.} Here we propose an alternative.\footnote{More accurately, a \emph{new} alternative. For discussion of other proposals see (Myrvold et al 2024).} We argue that the correlations in these experiments may be regarded as selection artefacts. As noted, selection bias often produces correlations between variables that are actually independent. It turns out that Bell correlations fit this model.

As we will explain, there are two ways to describe this conclusion. One is to say that the proposal \emph{avoids} nonlocality (without sacrificing realism). The other is to say that Bell nonlocality \emph{itself} turns out to be a selection artefact. This is a terminological issue, depending on what we mean by `nonlocality' when the new option is in view. We return to this question below (\S4.1), referring to Bell's own discussion of the term.

Either way, the upshot is that Bell correlations are less puzzling than they have seemed, in tension neither with relativity nor realism. `Less puzzling' does not imply `not puzzling at all'. The proposal identifies an escape route from some unwelcome consequences seemingly implied by Bell's work. It doesn't explain the novel features of QM on which the whole issue depends. Again, more on this below (\S4.2).

We begin with selection bias in general. Much of this material will be familiar to many readers, but we will introduce new terminology, at a couple of points.

\begin{figure}[htbp]
  \centering
  \includegraphics[width=2.38in]{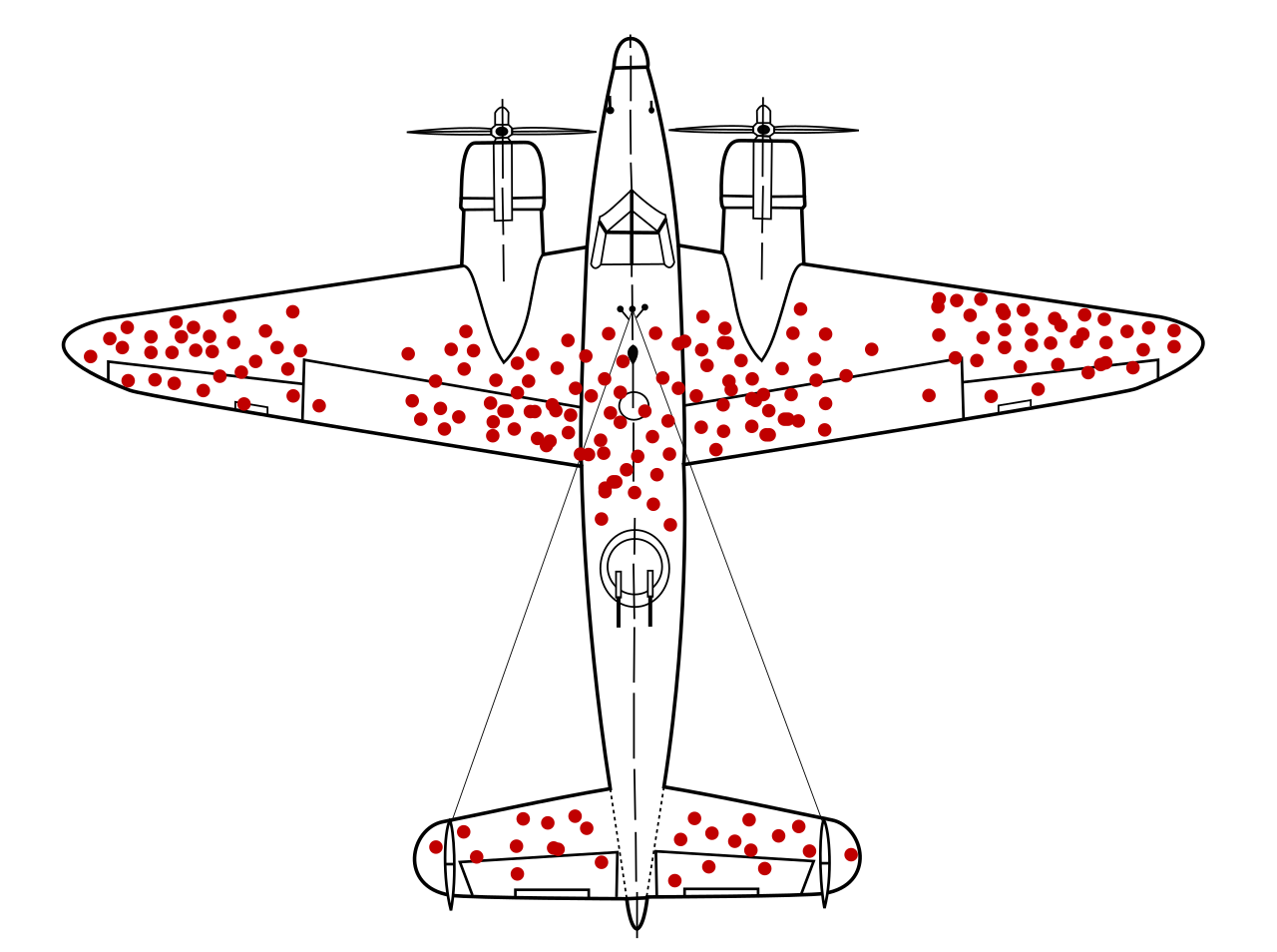}
  \caption{Survivorship bias.\protect\footnotemark}
  \label{fig:1}
\end{figure}
\footnotetext{Image by Granjean and McGeddon from
\url{https://en.wikipedia.org/wiki/Survivorship_bias}}

\section{What is selection bias?}

Readers may recognise Figure 1. It represents the distribution of damage found on bombers, returning from missions in WWII. (Think of the red dots as bullet holes.) Clearly, there is a strong correlation between location on the aircraft and the probability of damage. In some parts of the aircraft the probability is high, in other places close to zero.

We can imagine ways in which this correlation might be explained by causal dependencies. There are three main possibilities: (i) the geometry of the aircraft might be influencing the trajectory of the projectiles causing the damage; (ii) the projectiles might be influencing the position of the aircraft, the latter shifting itself to ensure that the damage is confined to these areas; (iii) some third variable -- a \emph{common cause} -- might be influencing the location of both the aircraft and the projectiles. None of these explanations is correct. As Abraham Wald pointed out,\footnote{The story may be apocryphal: \url{https://www.ams.org//publicoutreach/feature-column/fc-2016-06}.} the correlation results from the fact that we are looking at a biased subset of the data -- those aircraft that made it back to base. Aircraft with damage in other locations did not survive, and that's what's skewing the data, producing the correlation.

This is an example of selection bias. This particular variety is called \emph{survivorship bias,} for the obvious reason: the selection variable is `survival'. Survivorship bias is a subspecies of \emph{collider bias.} In the terminology of causal models, a collider is a variable influenced by two or more causes -- a node where two or more arrows `collide', in the graphical notation of direct acyclic graphs (DAGS; see Figure 2). Conditioning on a collider -- that is, selecting cases in which the collider variable takes a particular value -- typically produces a correlation between the contributing causes, even if they are in fact independent. In such cases, the \emph{apparent} dependence between the two causes is explained by the combination of two things: first, the \emph{actual} dependence between the two causes and their common effect; and second, the fact that the latter is being held fixed in a selection process. (Again, think of the bomber example.)

%\begin{figure}[htbp]
 % \centering
%  \includegraphics[width=2.49in]{figure02.png}
%  \caption{A collider}
%  \label{fig:2}
%\end{figure}

\begin{figure}
\centering
\begin{tikzpicture}
    % nodes set with absolute coordinates
    \node (a) at (0,0) {G};
    \node (F) at (3,3.3) {F};
    \node (b) at (6,0) {H};
\coordinate (coll) at (3,3); 
    % Directed edges
  \path[color=gray,rounded corners,line width=2pt] (a) edge (coll);
   \path[color=gray,rounded corners,line width=2pt] (b) edge (coll);

\end{tikzpicture}
\caption{A common effect, or collider} \label{fig:2}
\end{figure}

We propose here that Bell correlations are similar to collider bias -- indeed, some of them may \emph{be} collider bias. But it will be helpful to start further back, stressing first that not all selection bias is collider bias.

Why is selection bias not always collider bias? For two reasons, important in different ways. First, collider bias is a special case of something more general -- something not necessarily \emph{causal} at all. We will need the general notion below, so it will be helpful to clarify the point now. Second, there is another species of selection bias, whose effect is the opposite of collider bias. It \emph{masks} genuine dependencies rather than \emph{inducing} spurious dependencies. This kind of selection bias also has some important lessons for the quantum case. We will explain these two points in turn.

\subsection{Collider bias and its combinatorial cousins}

Collider bias typically induces correlations, but there are other cases of correlation-inducing selection bias, with no colliders involved. Most obviously, there are non-physical cases, e.g., in mathematics. Here, where variables are related \emph{logically} but not \emph{causally,} we find a similar kind of selection bias. For example, imagine a comparison of successive terms of the binary expansions of two irrational numbers, say $\pi$ and $e$. Let $\pi_{n}$ and $e_{n}$ be the $n$th terms in these expansions, respectively. For each $n$, there are four possibilities: the two terms $\pi_{n}$ and $e_{n}$ may be 00, 01, 10, or 11. Let's assume that these cases are equally likely -- that there's nothing that the value of $\pi_{n}$ tells us about that of $e_{n}$ and vice versa.\footnote{If that's not true about $\pi$ and $e$, we can choose different numbers.} Now let S be the set of natural numbers such that $\pi_{n}$ and $e_{n}$ are not both 1. Within S, $\pi_{n}$ and $e_{n}$ are highly correlated: if one is 1 then the other is 0. S behaves like a collider, but there is no causation in the picture.

It will be helpful to describe the underlying combinatorial structure of this kind of case. If A and B are two independent random binary variables, their joint distribution takes the uniform or `flat' form shown in the left-hand matrix in Figure 3.\footnote{We can imagine the two variables paired by position in a numerical ordering, as in the above example, or by timing, as in many physical examples.} The numbers in the cells represent the expected proportion of cases in each of the four possible combinations.

\begin{figure}[htbp]
  \centering
\begin{tikzpicture}%[remember picture,overlay]
%\begin{tikzpicture}[remember picture,overlay]
\filldraw[draw=black,fill=lightgray] (1,1) rectangle (2,2);
\filldraw[draw=black,fill=lightgray] (2,1) rectangle (3,2);
\filldraw[draw=black,fill=lightgray] (2,0) rectangle (3,1);
\filldraw[draw=black,fill=lightgray] (1,0) rectangle (2,1);

\coordinate[label=right:0] (0y) at (0.5,1.5);
\coordinate[label=right:1] (1y) at (0.5,0.5);
\coordinate[label=right:0] (0x) at (1.2,2.3);
\coordinate[label=right:1] (1x) at (2.2,2.3);

\coordinate[label=right:0.25] (2.5a) at (1.05,1.5);
\coordinate[label=right:0.25] (2.5b) at (2.05,1.5);
\coordinate[label=right:0.25] (2.5c) at (1.05,0.5);
\coordinate[label=right:0.25] (2.5d) at (2.05,0.5);

\coordinate[label=right:\textbf{B}] (Ay) at (0,1);
\coordinate[label=right:\textbf{A}] (Bx) at (1.7,2.7);

\filldraw[draw=black,fill=gray] (5,1) rectangle (6,2);
\filldraw[draw=black,fill=gray] (6,1) rectangle (7,2);
\filldraw[draw=black,fill=white] (6,0) rectangle (7,1);
\filldraw[draw=black,fill=gray] (5,0) rectangle (6,1);

\coordinate[label=right:0] (0y) at (4.5,1.5);
\coordinate[label=right:1] (1y) at (4.5,0.5);
\coordinate[label=right:0] (0x) at (5.2,2.3);
\coordinate[label=right:1] (1x) at (6.2,2.3);

\coordinate[label=right:\textbf{\textcolor{white}{0.33}}] (2.5a) at (5.0,1.5);
\coordinate[label=right:\textbf{\textcolor{white}{0.33}}] (2.5b) at (6.0,1.5);
\coordinate[label=right:\textbf{\textcolor{white}{0.33}}] (2.5c) at (5.0,0.5);
%\coordinate[label=right:0] (2.5d) at (2.05,0.5);

\coordinate[label=right:\textbf{B}] (Ay) at (4,1);
\coordinate[label=right:\textbf{A}] (Bx) at (5.7,2.7);

\end{tikzpicture}

  \caption{Correlator Bias -- a simple case}
  \label{fig:3}
\end{figure}

If some results are discarded -- e.g., as above, all cases where A=B=1 -- this produces a skewed distribution, as shown in the right-hand matrix. There are now dependencies between A and B -- if one of them is 1, the other must be 0. That's selection bias.

What is the relationship between this simple combinatorial model and collider bias? Remaining for simplicity with the binary case, it is that if a third variable D is causally influenced both by A and by B, then there will be a condition, expressible in terms of a distribution on this two-by-two matrix, for the cases in which D=0 and in which D=1. It may well be a probabilistic condition, but it can't be the flat distribution, uniform across the matrix. That would be the case in which D is \textbf{not} influenced by A and B. The right-hand matrix in Figure 3 corresponds to a case in which D takes one value if either A or B is 0, and the other value if A=B=1. Holding fixed the former case produces the distribution shown, with the dependence between A and B already described.

Colliders in causal models are thus one source of the kind of selection bias represented in Figure 3, but not the only source, as the mathematical case demonstrates. Even in the physical world, we may wish to model phenomena in non-causal terms -- probabilistic or combinatorial terms, for example. In such a model, holding fixed a variable may induce correlations between other variables, even though the model contains no causation and hence no colliders, strictly speaking. Finally, as we will see in a moment (\S2.4), this kind of selection bias can also arise at other points, even in cases that are represented in causal terms.

For all these reasons, the term \emph{collider bias} is unhelpfully restrictive. We will refer to this general form of selection bias as \emph{Correlator Bias}. A Correlator is a variable such that holding it fixed induces correlations between other variables. Colliders are typically Correlators, but Correlators need not be colliders.

\subsection{Selection bias from common causes}

We said above that there are two reasons why not all selection bias is collider bias. The second turns on a well-known property of common causes (Figure 4), whose effect on correlations is the opposite of a collider. The correlations produced by common causes \emph{disappear} if we `condition on the common cause' -- that is, if we select cases in which the common cause variable takes one particular value. The explanation is simple. If K co-varies with L because they both co-vary with a common cause J, then if we remove the variation of J, holding it fixed, we remove the co-variation between K and L.

%\begin{figure}[htbp]
%  \centering
 % \includegraphics[width=2.38in]{figure04.png}
 % \caption{A common cause}
 % \label{fig:4}
%\end{figure}

\begin{figure}
\centering
\begin{tikzpicture}
    % nodes set with absolute coordinates
    \node (a) at (0,0) {K};
    \node (F) at (3,-3.3) {J};
    \node (b) at (6,0) {L};
\coordinate (coll) at (3,-3); 
    % Directed edges
  \path[color=gray,rounded corners,line width=2pt] (coll) edge (a);
   \path[color=gray,rounded corners,line width=2pt] (coll) edge (b);

\end{tikzpicture}
\caption{A common cause} \label{fig:4}
\end{figure}

Here's an example. Suppose that K and L are rare diseases of mice. Both diseases are genetic disorders, caused by different abnormalities of the genes J that code for white hair colour. Few white mice get these rare diseases, but no grey mice do, because they lack the relevant genes. Imagine we don't know these facts, and we're interested in whether K and L are correlated. We examine many mice, looking for one or both conditions. If our lab breeds only white mice we will miss the correlation, because we are holding fixed the common cause J. This is selection bias, but here it is \emph{masking} a correlation, not producing one.

This kind of selection bias is called \emph{range restriction} (Dahlke \& Wiernik 2019). Unlike collider bias, it comes in two forms, depending on when the range is restricted. As described here, it is \emph{preselection bias.} The sample population is produced with the restriction built in. We could imagine the same error as \emph{postselection} bias. Suppose that the lab breeds mice of all the available colours, but that our assistant, Charlie, favours the white ones.\footnote{As Charlie puts it: `I dint know that mice were so smart. Maybe thats because Algernon is a white mouse. Maybe white mice are smarter than other mice.' Daniel Keyes, \emph{Flowers for Algernon} (Keyes 1959), p. 7.} Believing that we are kinder to mice than are other scientists in the lab, Charlie always selects white mice for our study. This masks the correlation between K and L in just the same way.

The fact that selection bias can arise from preselection will be important below. We are going to propose that the puzzling Bell correlations of QM are produced by selection bias. It will be Correlator Bias, not the correlation-masking kind of bias produced by holding fixed a common cause. But it is like the latter in that it typically involves \emph{preselection} -- a creation of a biased sample from scratch, as in the case where we breed only white mice.

\subsection{Correlators and Decorrelators}

More terminology. We introduced `Correlator' for the genus of which colliders are the most familiar species. Let's also introduce the term `Decorrelator', for cases that work like common causes -- cases in which fixing the variable in question masks the correlation between other variables. Hans Reichenbach called the common cause structure (Figure 4) a \emph{conjunctive fork} (Reichenbach 1956). We will call it a \emph{Decorrelating Fork,} to make its effect on correlations explicit. We will call the collider case a \emph{Correlating Fork.}

Figures 2 and 4 might suggest that there is a time-asymmetry between Correlators and Decorrelators -- that Correlating Forks have their central vertex in the future, while Decorrelating Forks have it in the past. This appearance is misleading, in both directions. With some care, we can find Decorrelating Forks (Reichenbach's conjunctive forks) with their vertex in the future (Arntzenius 1990, Weinberger 2026). And it is easy to find Correlating Forks with their vertex in the past -- \emph{In-the-Past Correlators} (IPCs), as we'll call them. There is a convenient example in the literature on QM and time-asymmetry.

\begin{figure}[htbp]
  \centering
  \includegraphics[width=4.3in]{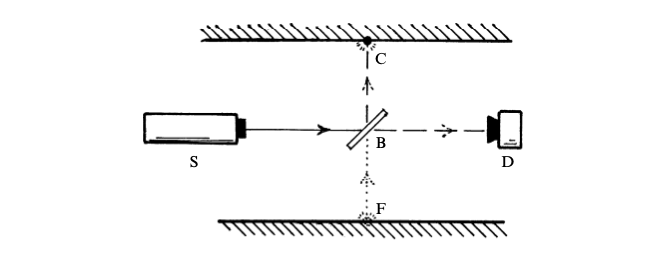}
  \caption{Penrose's time-asymmetry argument}
  \label{fig:5}
\end{figure}

\subsection{Penrose's In-the-Past Correlator}

Roger Penrose has long argued that there is a time-asymmetry in QM, in the component of the theory relevant to measurement. In (Penrose 1989) and (Penrose, 2004) he discusses the case shown in Figure 5, describing it as follows.

\begin{quote}
A source S randomly emits single high-energy photons \ldots\ aim-ed at a beam-splitter B, tilted at 45\textdegree\ to the beam. If transmitted through B, the photon activates a detector D (route SBD); if reflected, it is absorbed at the ceiling C (route SBC). The quantum squared-modulus rule correctly predicts probabilities \textonehalf, \textonehalf. On the other hand, given that D registers, the photon could have come from S (route SBD) or from the floor F (route FBD). Used in the reversed-time direction, the squared-modulus rule incorrectly retrodicts probabilities \textonehalf, \textonehalf, which should be 1, 0. {(}Penrose 2004, 820{)}
\end{quote}

In response, many critics have pointed out that the apparent time-asymmetry results from the fact that Penrose is holding fixed an initial condition (see, e.g., Baez 1996). If we add a second, equally probable source of photons on the floor at F, the asymmetry disappears. In this modified version, the source location (S or F) acts as an IPC. Holding it fixed to either value produces Penrose's time-asymmetry. Specifically, it ensures that any two photons arriving at D are correlated in having the same origin. Allowing sources at S and F to be equally likely removes this correlation, along with the time-asymmetry.

This example has several helpful lessons, for our present purposes. It confirms the existence of IPCs, showing that even in the causal world, Correlators need not be colliders. The causal structure of Penrose's QM case might conceivably be regarded as unclear, but it is easy to construct classical analogues. There, it will be obvious that the location of the source is not a collider. The case also demonstrates that preselection on an IPC may explain a real-world correlation -- even a puzzling correlation. And it shows that it is easy to miss this simple explanation, and hence overlook a solution to the puzzle concerned.

Classical IPCs are exceedingly common,\footnote{Arguably all imposition of low entropy initial conditions has this character, including the so-called Past Hypothesis itself.} and Penrose's case illustrates that similar cases arise in QM. However, the present proposal rests on the fact that IPCs also arise in a \emph{new} way in QM, without classical analogues. The core of the proposal is that the entangled states prepared at the beginning of ordinary two-particle Bell experiments are themselves IPCs of this novel variety. We will return below (\S5.3) to the respects in which they differ from the example just described, but the proposal has the following in common with Penrose's case. It claims that the observed correlations between the two sides of a Bell experiment are the result of holding fixed an IPC, as we do when we make repeated runs of the experiment with the same initial state.

\subsection{A Minimal Selection Bias Criterion}

Before we turn to QM, let's formulate a criterion for selection bias. What does such bias require? At a minimum, presumably, a distinction between a larger `super-ensemble' of cases and a smaller `sub-ensemble', the latter picked out of the former by some sort of selection process.\footnote{What is a selection process? A good question, which we defer, though the cases we need will not be controversial.} Selection bias requires correlations that \emph{differ} between the super-ensemble and the sub-ensemble. Call this the \emph{Minimal Selection Bias Criterion} (MSBC).

We've seen that the difference required by MSBC may go in either direction. Selection may \emph{remove} as well as \emph{produce} correlations. We've also seen that a super-ensemble need not be an ensemble of \emph{actual} cases. When preselection is involved, the selection may be one that prevents some of the super-ensemble from coming into existence in the first place. In such cases, much of the super-ensemble consists of possible but unrealised cases -- it is partly a \emph{virtual} super-ensemble, as we might say. This will be important in the QM case: ordinary preparation of the initial state of a Bell experiment is a preselection of this kind.\footnote{This is also true of Penrose's case, of course. In the original version as he imagines it, there is no option for an \emph{actual} source at F.}

\begin{figure}[t]
\centering
\begin{tikzpicture}
      %\node[state,fill=green] (a) at (0,0) {$a$};
    \node[draw,rectangle,minimum width=0.6cm,minimum height=0.5cm] (I) at (3.1,-2.5) {\textbf{\textsf{C}}};
 \node[] (C) at (3.1,-3.3) {$C$};

    \node[draw,rectangle,minimum width=0.6cm,minimum height=0.5cm] (Abox) at (1.2,0) {\textbf{\textsf{A}}};
   \node[] (Alabel) at (1.2,1.7) {$A$};
    \node[] (A0) at (0.8,1) {0};
    \node[] (A1) at (1.6,1) {1};
    \node[] (a) at (0.5,-1) {$a$ \footnotesize{(0 or 1)}};
    \node[draw,rectangle,minimum width=0.6cm,minimum height=0.5cm] (Bbox) at (5,0) {\textbf{\textsf{B}}};
     \node[] (Blabel) at (5,1.7) {$B$};
       \node[] (B0) at (4.6,1) {0};
    \node[] (B1) at (5.4,1) {1};
    \node[] (b) at (5.7,-1) {$b$ \footnotesize{(0 or 1)}};
 %   \node[state,fill=green] (b) at (6.2,0) {$b$};

   \path [dashed,-] (Abox) edge (A0);
   \path [dashed,-] (Abox) edge (A1);
   \path [->] (a) edge (Abox.south west);
    \path [dashed,-] (Bbox) edge (B0);
   \path [dashed,-] (Bbox) edge (B1);
    \path [->] (b) edge (Bbox.south east);

   \draw [color=blue,line width=1.5pt,-] (I.north) -- (Abox.south east);
   \draw [color=blue,line width=1.5pt,-] (I.north) -- (Bbox.south west);
\end{tikzpicture}
\caption{A two-particle Bell experiment} \label{fig:6}
\end{figure}

\section{Bell experiments}

\subsection{Introducing Factorizability}

Figure 6 depicts a typical two-particle Bell experiment. Pairs of entangled particles are prepared at \textbf{C} in an initial state \emph{C.} In one common version of the experiment, the particles are photons, prepared in the so-called \emph{singlet state.} The particles are directed to experimenters at \textbf{A} and \textbf{B}, who each choose a setting for their measurement device, and record an outcome. In the photon version, the measurement devices are polarizers. The settings and outcomes both take two possible values, represented as 0 and 1. We use \emph{a} and \emph{b} as variables for the settings at \textbf{A} and \textbf{B}\emph{,} respectively, and \emph{A} and \emph{B} for the outcomes. QM allows us to calculate the joint probabilities of the four possible combinations of outcomes, given the initial state \emph{C} and the four possible combinations of settings. We write these probabilities as $P(A,B \mid a,b,C)$.

The intuitive idea of Bell's condition of locality is that when the experiments at \textbf{A} and \textbf{B} are widely-separated -- in particular, if they are spacelike separated in the sense of special relativity -- then the probabilities at \textbf{A} and \textbf{B} should be independent, for a given \emph{C}. The probability for an outcome at \textbf{A} should not depend on the setting or outcome at \textbf{B}, and vice versa. If this is true, then the probabilities are said to be \emph{Factorizable:}

\begin{equation}\tag{Factorizability}\label{eq:fact}
  P(A,B \mid a,b,C) = P(A \mid a,C)\,P(B \mid b,C)
\end{equation}

Bell showed that Factorizability implies a numerical constraint on the expected values of such experiments, now known as a \emph{Bell inequality.} Bell saw that QM predicts that such constraints are violated in some cases, predictions now confirmed in many experiments. In most cases, the actual Bell inequality used is the so-called Clauser-Horne-Shimony-Holt (CHSH) Inequality, derived in a refinement of Bell's argument (Clauser et al 1969).

Experimental violations of Bell inequalities are the basis for the claim that Factorizability fails in QM, and hence that the quantum world is nonlocal. However, we will now argue that failure of Factorizability is a selection artefact, requiring no spacelike dependency of any other sort. We will approach this conclusion in several steps.\footnote{Does this mean that failure of Factorizability is wrongly equated with nonlocality, or that so-called nonlocality may itself be a selection artefact? As noted in \S1, this becomes a terminological choice, to which we return below.}

%\begin{figure}[htbp]
%  \centering
 % \includegraphics[width=2.32in]{figure07.png}
 % \caption{Bell experiment with four possible initial Bell states}
 % \label{fig:7}
%\end{figure}

\begin{figure}[t]
\centering
\begin{tikzpicture}
      %\node[state,fill=green] (a) at (0,0) {$a$};
    \node[draw,rectangle,minimum width=0.6cm,minimum height=0.5cm] (I) at (3.1,-2.5) {\textbf{\textsf{C}}};
\node[] (I0) at (2.2,-3.4) {$C_0$};
 \node[] (I1) at (2.8,-3.4) {$C_1$};
 \node[] (I2) at (3.4,-3.4) {$C_2$};
 \node[] (I3) at (4,-3.4) {$C_3$};
 %\node[] (Prep) at (2.2,-4.7) {\textbf{Preparation}};
 \node[] (C) at (3.1,-4) {$C$};

  \path [dashed,-] (I1) edge (I);
  \path [dashed,-] (I0) edge (I);
  \path [dashed,-] (I2) edge (I);
  \path [dashed,-] (I3) edge (I);
 % \path [->] (Prep) edge (I0);

    \node[draw,rectangle,minimum width=0.6cm,minimum height=0.5cm] (Abox) at (1.2,0) {\textbf{\textsf{A}}};
   \node[] (Alabel) at (1.2,1.7) {$A$};
    \node[] (A0) at (0.8,1) {0};
    \node[] (A1) at (1.6,1) {1};
    \node[] (a) at (0.5,-1) {$a$};
    \node[draw,rectangle,minimum width=0.6cm,minimum height=0.5cm] (Bbox) at (5,0) {\textbf{\textsf{B}}};
     \node[] (Blabel) at (5,1.7) {$B$};
       \node[] (B0) at (4.6,1) {0};
    \node[] (B1) at (5.4,1) {1};
    \node[] (b) at (5.7,-1) {$b$};
 %   \node[state,fill=green] (b) at (6.2,0) {$b$};

   \path [dashed,-] (Abox) edge (A0);
   \path [dashed,-] (Abox) edge (A1);
   \path [->] (a) edge (Abox.south west);
    \path [dashed,-] (Bbox) edge (B0);
   \path [dashed,-] (Bbox) edge (B1);
    \path [->] (b) edge (Bbox.south east);

   \draw [color=blue,line width=1.5pt,-] (I.north) -- (Abox.south east);
   \draw [color=blue,line width=1.5pt,-] (I.north) -- (Bbox.south west);

\end{tikzpicture}
\caption{Bell experiment with four possible initial Bell states} \label{fig:7}
\end{figure}

\subsection{Bell correlations by preselection: a toy model}

Bell experiments with the geometry of Figure 6 can be performed with any of the four so-called Bell states as the initial preparation. Let's write these four states as $C_{0}, C_{1}, C_{2}, C_{3}$, where $C_{0}$ is the singlet state. (See Figure 7.) For each of these states, there is a probability distribution $P(A,B \mid a,b,C_{i})$ giving the predicted joint probability of outcomes (\emph{A,B}) given settings (\emph{a,b}) and initial state $C_{i}$.

Let's imagine a version of Figure 7 in which the initial Bell state is chosen at random on each run, with equal probability for each of the four possibilities. This device generates an ensemble of results in which (i) there are no Bell inequality-violating correlations in the ensemble as a whole, but (ii) there are such correlations in each of the four sub-ensembles defined by the initial state $C_{i}$. Factorizability holds in the full ensemble -- in other words, $P(A,B \mid a,b) = P(A \mid a)\,P(B \mid b)$ there -- but fails in each of the sub-ensembles.

On the face of it, then, we have met the MSBC test for treating \emph{some} Bell correlations as selection artefacts, in a real QM case. In a case in which selection produces correlations, MSBC calls for two things: an uncorrelated super-ensemble, and a selection procedure that picks out correlated sub-ensembles of this super-ensemble. In the case just described, this selection is performed by the random device that controls the initial state $C_{i}$. Bell correlations emerge in the sub-ensembles defined by a fixed value of $C_{i}$. The initial state $C_{i}$ is therefore a Correlator, in the sense defined above.

Thus we have a toy model using real QM components, in which the Bell inequality-violating correlations between the two sides of the experiment can be represented as selection artefacts. If we could argue that the toy model is actually \emph{general} -- a good representation of \emph{arbitrary} Bell tests with these components -- then we would have something interesting. We'd have an argument that such Bell correlations are selection artefacts.

\subsection{The A to Z of generalisation}

How does our toy model differ from ordinary, real-world Bell tests? For one thing, it combines four kinds of Bell experiment in a single apparatus. But this difference is superficial. With some coordination, we could build the device from four single-purpose Bell tests of the usual kind. The central randomiser might send an instruction to one of four experimenters, each conducting an experiment with a single initial Bell state: Professors A in Paris, B at CERN, C at Berkeley, and Z in Vienna, perhaps. Each of the professors agrees to trigger a run of their apparatus when and only when they receive an instruction from Central Command (the randomiser). Otherwise, what they do in their individual laboratories is entirely as normal. But the representation of their individual, Bell-inequality-violating results as selection artefacts is exactly as in the original case.

Does this mean that we have successfully generalised the toy model, and are free to regard the results of ordinary, uncoordinated instances of these Bell tests as selection artefacts; and if not, why not? Perhaps the barrier to generalisation is that some of the professors are unreliable -- much of the time, perhaps, at most one of our professors is actually in the lab, ready to run an experiment. But this can't be the crucial point. We saw that preselection bias may involve \emph{virtual} super-ensembles. In the experiment just described, it makes no difference if the telegraph to three out of the four labs is broken, or indeed if three labs have suddenly ceased to exist. We still have our virtual ensemble, allowing us to apply the diagnosis of selection bias to the results in the remaining lab.

The real barrier to generalisation lies in the assumptions we made about the randomiser -- namely, that each of the four initial Bell states is \emph{equally likely.} Without that, some form of failure of Factorizability will survive in the full ensemble. In other words, our uncorrelated super-ensemble was constructed artificially, by imposing equiprobable choice of the initial Bell state. Why not regard this as gerrymandering, rather than a construction with application in the real world?

\subsection{Experimental control and the difference between past and future}

We are proposing that in the kind of experiment depicted in Figure 6, Bell correlations between \textbf{A} and \textbf{B} arise because we fix a Correlator at \textbf{C} when we prepare the initial state. Putting the choice of the initial state in the hands of a random device reduces our degree of control, but it doesn't remove it altogether. The probabilities of the four possible initial states are still controlled by us -- we built the randomiser, after all, and we could have dialed in any distribution we chose of the probabilities of the four possible initial states.

What would the probabilities be if there were \emph{no experimental control whatsoever?} If the four initial states were equally likely in that case, we could take the case with the randomiser to be a model of Nature's own probabilities. Readers may find this question perplexing. Of course we control the initial states of our experiments, at least in principle! How could it make sense to throw that away? But consider this. Whatever this initial control of experiments amounts to, it is time-asymmetric. We don't control the final states of experiments in the same way. This should give us pause, if we are inclined to insist that initial control is something fundamental, something whose absence doesn't make sense. If that were true, it would amount to a fundamental time-asymmetry.

Yet it is widely believed that fundamental physics, including quantum theory, is time-symmetric.\footnote{Or CPT-symmetric, if we want to be careful.} Observed time-asymmetries are standardly attributed to a boundary condition in the early universe, imposing very low entropy at that point. If this story is correct, then our own existence, and our time-asymmetric ability to control our environments, is also attributable to this low entropy past (no doubt somewhat remotely).

This standard picture provides the resources, in principle, to consider the case without initial control. Whether in classical or quantum versions, the picture involves a time-symmetric measure of possible histories for the universe, from which our observed history emerges by imposing the low entropy past. All manner of unlikely possibilities may be considered in an imaginary regime lacking the low entropy constraint. Random pieces of quantum optics will be an easy trick, compared to Boltzmann Brains (Carroll 2017).

For present purposes, the only point we need is that this `unconstrained' case, in which the initial state of the Bell experiment is not under experimental control, is not illicit in some deep theoretical sense. The question about the natural measure on the space of unconstrained possibilities at \textbf{C} is a respectable one, albeit an unusual one.

From this point, there are two paths our argument might take. First, we could simply \emph{postulate} the measure we want. In other words, we could postulate that the example with a randomiser is an adequate model of the unconstrained case. This postulate comes at very modest cost -- why should we assume that the measure is anything else? -- and it has a generous payoff. It ensures that the Bell correlations in experiments with the form of Figure 6 are entirely explicable as selection artefacts, emerging from an uncorrelated super-ensemble when we fix the value of the Correlator at \textbf{C}. The proposal could then be regarded as an hypothesis, relying on this assumption about the measure.\footnote{Readers may object that the postulate is unnecessary. If the Universe favoured some other measure in the unconstrained case, then the proposal could proceed as before, except that some of the relevant preselection would be being done by the Universe itself. We set this suggestion aside here.}

The second strategy makes a virtue of the ordinary asymmetry between past and future. As we just observed, we normally have control of the initial but not the final conditions of our experiments. We have been interested in the case in which this initial control is imagined to be absent. Could we find a model for that case in a version of our experiment in which the relevant variable sits in the future, rather than the past? In the future it would be uncontrolled anyway, with no act of imagination needed. This thought leads us to a different category of Bell experiments, which also happen to be interesting for other reasons.

%\begin{figure}[htbp]
%  \centering
 % \includegraphics[width=2.82in]{figure08.png}
%  \caption{A W-shaped Bell experiment with entanglement swapping}
 % \label{fig:8}
%\end{figure}

\subsection{Bell correlations by postselection}

Figure 8 depicts an important experimental protocol, used for some of the best tests of Bell correlations. We can call it the W-shaped protocol, referring to the shape of its arrangement in spacetime. It relies on so-called \emph{entanglement swapping.} Two pairs of entangled particles are produced independently, one pair at the source \textbf{S\textsubscript{1}} and the other at the source \textbf{S\textsubscript{2}}. One particle from each pair is sent to a measurement device at \textbf{M}, which performs a Bell state measurement on the two particles -- a measurement with four possible outcomes, shown in Figure 8 as $M_{0}, M_{1}, M_{2}$ and $M_{3}$.

\begin{figure}[t]
\centering
\begin{tikzpicture}
    \node[draw,rectangle,minimum width=0.6cm,minimum height=0.5cm] (I1) at (3,-2.5) {\textbf{\textsf{S$_1$}}};
    \node[draw,rectangle,minimum width=0.6cm,minimum height=0.5cm] (I2) at (7,-2.5) {\textbf{\textsf{S$_2$}}};
    \node[draw,rectangle,minimum width=0.8cm,minimum height=0.5cm] (W) at (5,3) {\textbf{\textsf{M}}};
\node[] (W0) at (3.8,4) {$M_0$};
\node[] (W1) at (4.6,4) {$M_1$};
 \node[] (W2) at (5.4,4) {$M_2$};
 \node[] (W3) at (6.2,4) {$M_3$};
%\node[] (Out) at (5,4.7) {$M$};
 %\node[] (I10) at (3,-3.6) {\textbf{0}};
  %\node[] (I20) at (7,-3.6) {\textbf{0}};

 \path [dashed,-] (W1) edge (W);
 \path [line width=1.5pt,-] (W0) edge (W);
\path [dashed,-] (W2) edge (W);
 \path [dashed,-] (W3) edge (W);
%\path [->] (I10) edge (I1);
%\path [->] (I20) edge (I2);

    \node[draw,rectangle,minimum width=0.6cm,minimum height=0.5cm] (Abox) at (1,0) {\textbf{\textsf{A}}};
  \node[] (Alabel) at (1,1.7) {$A$};
    \node[] (A0) at (0.6,1) {0};
    \node[] (A1) at (1.4,1) {1};
    \node[] (a) at (0.3,-1) {$a$};
    \node[draw,rectangle,minimum width=0.6cm,minimum height=0.5cm] (Bbox) at (9,0) {\textbf{\textsf{B}}};
      \node[] (Blabel) at (9,1.7) {$B$};
       \node[] (B0) at (8.6,1) {0};
    \node[] (B1) at (9.4,1) {1};
     \node[] (b) at (9.7,-1) {$b$};
       \node[] (bc) at (4.4,5.2) {\footnotesize{\textbf{Postselection}}};
       \path [<-] (bc) edge (W0);

   \path [dashed,-] (Abox) edge (A0);
   \path [dashed,-] (Abox) edge (A1);
    \path [->] (a) edge (Abox.south west);
    \path [dashed,-] (Bbox) edge (B0);
   \path [dashed,-] (Bbox) edge (B1);
   \path [->] (b) edge (Bbox.south east);
      \draw [color=blue,line width=1.5pt,-] (I1.north) -- (Abox.south east);
   \draw [color=blue,line width=1.5pt,-] (I2.north) -- (Bbox.south west);
      \draw [color=blue,line width=1.5pt,-] (I1.north) -- (W.south);
       \draw [color=blue,line width=1.5pt,-] (I2.north) -- (W.south);

  % \node[circle,minimum size=0.7cm,color=gray,draw] (i) at (0,0){\footnotesize{I}};
  %  \node[circle,minimum size=0.7cm,color=gray,draw] (ii) at (5,1.4){\footnotesize{II}};
    % \node[circle,minimum size=0.7cm,color=gray,draw] (iii) at (10,0){\footnotesize{III}};
    %  \node[circle,minimum size=0.7cm,color=gray,draw] (iv) at (2,-2.5){\footnotesize{IV}};
     %  \node[circle,minimum size=0.7cm,color=gray,draw] (v) at (8,-2.5){\footnotesize{V}};
  
\end{tikzpicture}
\caption{A W-shaped Bell experiment with entanglement swapping} \label{fig:8}
\end{figure}

The remaining two particles go to measuring devices at \textbf{A} and \textbf{B}, respectively. The experimenters each choose a setting and record an outcome, in the usual way. The expected results are as follows. There are no Bell inequality-violating correlations between \textbf{A} and \textbf{B} in the ensemble of results as a whole, but such correlations emerge if we postselect on any of the four outcomes at \textbf{M} -- e.g., on $M_{0}$, as shown in Figure 8. As noted, some of the best real Bell experiments use this protocol.

It is easy to see that the case meets our MSBC test for the presence of selection artefacts. The super-ensemble is the set of results as a whole, and the sub-ensembles are the four sets of results we obtain by postselection on a single value of \emph{M} at \textbf{M}. In the sense implied by MSBC, Bell correlations in such experiments simply \emph{are} selection artefacts -- there's no question about the matter. Because the selection is inducing rather than masking correlations, the output variable \emph{M} is a Correlator, in our new terminology.

Two clarifications. First, the diagnosis just given applies to the Bell correlations `across the W' between \textbf{A} and \textbf{B}. It says nothing, as it stands, about the entanglement in the two wings of the experiment, needed to produce the results at \textbf{M} that in turn yield postselected correlations between \textbf{A} and \textbf{B}. We set this point aside here, with the following note. If we have a diagnosis that works in ordinary two-particle (`V-shaped') Bell experiments, then we could expect to apply it here, to deal with the wings as well. The W experiments would then be hybrid cases, involving selection at all three vertices: preselection at \textbf{S\textsubscript{1}} and \textbf{S\textsubscript{2}}, and postselection at \textbf{M}.

Second, our diagnosis said nothing about a significant feature of the geometry of these W-shaped experiments, that makes a big difference to how they are represented in orthodox QM. The measurement at \textbf{M} can have three possible locations with respect to those at \textbf{A} and \textbf{B}.\footnote{We are ignoring some hybrid possibilities, e.g., cases in which \textbf{A} and \textbf{B} are not spacelike separated from one another.} It may be in their absolute future -- so-called \emph{delayed choice} entanglement swapping (DCES) -- in their absolute past, or somewhere in between. In the orthodox QM picture with collapse, this makes a big difference to the description of the case. In the DCES case the measurements at \textbf{A} and \textbf{B} collapse the state of the second particle in each pair, influencing the states reaching \textbf{M}. In the absolute past case, the influence goes the other way: the measurement at \textbf{M} affects the states of the particles received at \textbf{A} and \textbf{B}.

In the DCES case, as several authors have noted, this means that we have something like a causal story, for how information at \textbf{A} and \textbf{B} can make a difference to the outcome at \textbf{M}. It becomes plausible to regard \textbf{M} as a collider, in other words, and hence to treat the resulting \textbf{A}--\textbf{B} correlations as a case of collider bias.\footnote{See (Gaasbeek 2010), (Egg 2013), (Fankhauser 2019), (Bacciagaluppi \& Hermens 2021), (Price \& Wharton 2021), and (Mjelva 2024).} But this story won't work, apparently, in the other possible W geometries.

However, we stressed earlier that not all Correlators are colliders. If we were trying to force Bell correlations into a box labeled \emph{Collider Bias} then the geometry of the W cases would certainly be a concern. But MSBC doesn't mention causal structure, merely statistical structure:  a difference in the correlations in the super-ensemble and selected sub-ensembles. By this purely statistical criterion, it seems indisputable that the Bell correlations are selection artefacts, associated with a Correlator at \textbf{M}, in all the W geometries. We need to postselect on the result at \textbf{M}, wherever \textbf{M} stands in spacetime with respect to \textbf{A} and \textbf{B}.

In sum, our model applies easily to the W cases -- including, as we said, to some of the best recent Bell tests. To take us back to V cases, note that QM itself provides a flat distribution over the four possible outcomes at \textbf{M} -- all four possibilities have equal probability. This seems a promising analogue for the corresponding probabilities at \textbf{C}, in the two-particle case we were discussing in \S3.4 -- the imaginary case in which no distribution of probabilities for inputs at \textbf{C} is supplied by the experimenter.

In principle, these experiments could be conducted so that Alice and Bob, confined in their laboratories, do not know whether the experiment as a whole has the V-shaped geometry of Figure 6 or the W-shaped geometry of Figure 8. The two experiments can be configured so that the joint probability of outcomes (\emph{A,B}) given settings (\emph{a,b}) and outcome $M_{i}$ at \textbf{M} in the W case, is identical to the joint probability of outcomes (\emph{A,B}) given settings (\emph{a,b}) and initial state $C_{i}$ at \textbf{C} in the V case -- so that $P(A,B \mid a,b,M_{i}) = P(A,B \mid a,b,C_{i})$, in other words. In this case, it is as though the vertex at \textbf{C} has simply been `flipped' into the future, keeping the probabilities the same, but changing inputs to outputs.

\subsection{Summary}

We have proposed that Bell inequality-violating correlations be regarded as selection artefacts, the relevant selection being the choice of the initial state at \textbf{C} in the V-shaped case of Figure 6, or the output state at \textbf{M} in the W-shaped case in Figure 8. In either case, this meets the MSBC test, an uncorrelated super-ensemble being readily at hand in both cases.\footnote{Does such a super-ensemble need to be unique, where it is partly virtual? Our mice suggest not. The correlation due to the common cause will be visible if we add to our sample grey mice, black mice, or both, for example.}

In the V-shaped case the super-ensemble may be largely a \emph{virtual} ensemble, in the sense that many of its members are never brought into existence. However, we noted that this is common not only in cases where the source of selection bias is a common cause (remember our mice), but also with IPCs such as Penrose's experiment. In this respect, the appeal to virtual ensembles introduces nothing new. Just as with the white mice, the artefacts produced by selection in Figure 6 arise equally from preselection or postselection. We can choose to run the experiment so that it is locked to a single initial state at \textbf{C}, or we can take what comes, and throw away the cases we don't want. Either way, the resulting Bell correlations meet the MSBC condition to count as selection artefacts.

\section{Discussion}

\subsection{Nonlocality -- diagnosis or cure?}

Now to a question deferred at the beginning. Should we regard this proposal as a way of \emph{explaining} Bell nonlocality, or as a way of \emph{avoiding it altogether?} This is a subtle matter, for it depends on what we mean by `locality'. We can explain this point with reference to Bell's own discussion of the term. Bell's most extended treatment of locality is in his late piece `La nouvelle cuisine' (Bell 1990), written a few months before his untimely death. He begins there with what he calls the \emph{Principle of Local Causality} (PLC).

\begin{quote}
The direct causes (and effects) of events are near by, and even the indirect causes (and effects) are no further away than permitted by the velocity of light. (Bell 1990, 239)
\end{quote}
He then says that this `principle of local causality is not yet sufficiently sharp and clean for mathematics', and follows this immediately with a warning:

\begin{quote}
Now it is precisely in cleaning up intuitive ideas for mathematics that one is likely to throw out the baby with the bathwater. So the next step should be viewed with the utmost suspicion. (1990, 239)
\end{quote}

As we'll see, Bell was right to be cautious, for he himself misses an important baby -- it is selection bias. After this warning, Bell formulates a second principle of local causality, motivated by simple considerations of light cone structure, and the like. It is very close to Factorizability itself, which Bell takes to be an immediate consequence. We don't need the details of the second principle here,\footnote{For the details, see the discussion in (Myrvold et al 2024), to which we are indebted here.} but it is worth noting that Bell makes a virtue of the shift to talk of probability, rather than cause and effect:

\begin{quote}
Note, by the way, that our definition of locally causal theories, although motivated by talk of ``cause'' and ``effect'', does not in the end explicitly involve these rather vague notions. (1990, 240)
\end{quote}

We'll return to Bell's cautiousness about causation below (\S4.2). When Bell introduces Factorizability explicitly, he says that it is `consequent on local causality' (1990, 244), meaning his second version of the latter. For our purposes, it will be sufficient to restate Factorizability as we had it earlier, in a slightly simpler form than Bell's.%: $P(A,B \mid a,b,C) = P(A \mid a,C)\,P(B \mid b,C)$.
\footnote{The main modification we have made to Bell's formulation is to remove possible `hidden variables' $\lambda$ from the conditions of all these conditional probabilities. Our present framework is purely operational, and this simplification makes no difference to the conclusions drawn below.} 

\begin{equation}\tag{Factorizability}\label{eq:fact2}
  P(A,B \mid a,b,C) = P(A \mid a,C)\,P(B \mid b,C)
\end{equation}

\noindent About this, Bell says the following (we have changed the notation slightly to match our own):

\begin{quote}
Now this formulation has a very simple interpretation. It exhibits \emph{A} and \emph{B} as having no dependence on one another, nor on the settings of the remote polarizers (\emph{b} and \emph{a} respectively), but only on the local polarizers (\emph{a} and \emph{b} respectively) and on the past causes, {[}\emph{C}{]} \ldots{} We can clearly refer to correlations which permit such factorization as `locally explicable\textquotesingle. Very often such factorizability is taken as the starting point of the analysis. Here we have preferred to see it not as the \emph{formulation} of `local causality\textquotesingle, but as a consequence thereof. (1990, 243)
\end{quote}

It should now be clear that we have two options. The first is to take the usual path, \emph{defining} locality as Factorizability. In this case, the present proposal implies that locality fails, because Factorizability fails. But the failure of Factorizability is a selection artefact, rather than an indication of any direct causal influence from \textbf{A} to \textbf{B}. It is a benign form of nonlocality, in other words, in no tension with relativity or realism.

Once we see this, we may prefer the second option: to follow Bell, distinguishing Local Causality and Factorizability. The present proposal then shows that Bell was wrong that the latter is an automatic consequence of the former. Factorizability fails, though Local Causality does not. Bell was right to be suspicious of the step from one to the other. The baby discarded with the bathwater is the possibility of selection artefacts.

As we said, the choice between these two options is a terminological matter. The need to make this choice should not obscure what the two options have in common. In either version, the present proposal avoids the seeming tension between Bell correlations, on the one hand, and the conjunction of relativity and realism, on the other.\footnote{As we explain in \S5 below, it turns out that we may need to allow a third option. Depending on what we choose to call the connections established by constrained Correlators, we may wish to say that even PLC fails -- though again, in a way that poses no challenge to relativity or realism.}

\subsection{Limitations and mechanisms}

Now to some limitations of the proposal. We noted that although it offers an escape from Bell's apparent tension between realism and relativity, it doesn't explain how the problem arises in the first place. There are several points to emphasise here.

The proposal rests on the observation that the results of a typical Bell experiment with fixed initial state (Figure 6) may be regarded as a sub-ensemble of the results of a broader experiment, with variable initial state (Figure 7). The \textbf{A}--\textbf{B} correlations in the sub-ensemble may thus be interpreted as selection artefacts, in the way described. However, the proposal makes no claim to explain the correlations in the super-ensemble on which the representation depends. These broader \textbf{A}--\textbf{B}--\textbf{C} correlations are simply taken for granted.

These broader relations are standard operational correlations of QM. So they are available off the shelf, and the burden of explaining them does not fall on the present proposal alone. However, not all possible explanations will embrace the claimed advantage of the proposal, that of avoiding direct \textbf{A}--\textbf{B} dependence. The de Broglie-Bohm view postulates just such a connection, for example. By the lights of our proposal that seems an unnecessary cost, but a cost is not a decisive counter argument. The proposal is not \emph{incompatible} with the de Broglie-Bohm view, or with other QM models with direct spacelike causation.

If we are interested in mechanisms for the operational correlations that take advantage of the proposal, where should we look? A natural approach is to invoke the framework of causal models, thought of as an abstract first step on an intended path to something more concrete. We offer some recommendations for such an approach, drawing on the present discussion and some of Bell's own remarks about causation in physics.

Our proposal suggests a question for any application of causal models to the cases we have discussed. In the familiar, classical, physical world, many Correlators are also colliders, with the temporal orientation of Figure 2. We noted that in the DCES version of the W case in Figure 8, the central measurement result at \textbf{M} also looks like a collider, in orthodox QM -- we have what looks like a causal story, via collapse.\footnote{In DCES cases the most straightforward causal story treats outcomes as well as settings at \textbf{A} and \textbf{B} as contributing causes at \textbf{M}. We set this aside here.} Hence the question: should the other W cases and the V case \emph{also} be modelled as involving colliders, if we are interested in underpinning the operational proposal with a causal model?

If we answer `Yes' we will be envisaging \emph{retro}causality in some cases. This idea has often been suggested in QM,\footnote{For an overview, see (Friedrich \& Evans 2023), (Wharton \& Argaman 2020).} and it may be playing a role here. This point connects with Bell's own caution about the use of the term causation in physics, mentioned above. In an early piece he expresses one of his reservations, and his view of its resolution, as follows:.

\begin{quote}
In this matter of causality it is a great inconvenience that the real world is given to us once only. We cannot know what would have happened if something had been different. We cannot repeat an experiment changing just one variable; the hands of the clock will have moved, and the moons of Jupiter. Physical theories are more amenable in this respect. We can \emph{calculate} the consequences of changing \textbf{free} elements in a theory, \textbf{be they only initial conditions}\emph{,} and so can explore the causal structure of the theory. (Bell 1977, 101; emphasis added)
\end{quote}

Bell puts a time-asymmetry in by hand here, with the phrase in bold. He himself doesn't seem to have been inclined to investigate the source of this time-asymmetry. As noted above (\S3.4), a standard view associates it with the presence of a low entropy boundary condition, in the early history of the universe.

If we wish to discuss causality in an imaginary regime without such a boundary constraint, we can remove the time-asymmetry that Bell puts in by hand, by deleting the emphasised phrase. We can calculate the consequences of changing other elements, and be liberal about what we count as a free or exogenous element. With these tweaks to Bell's own conception of causality in physics, we get a time-symmetric picture.\footnote{The Future-Input-Dependent models of (Wharton \& Argaman 2020) are a step in this direction, but do not envisage treating outcomes as contributing causes, in the sense mentioned in fn. 20.} In this picture, we get colliders where we want them, in all the experimental geometries.

Some readers may feel that the resulting time-symmetric conception of causality is too thin to deserve the name. Why not simply speak of statistical dependence? This is largely a terminological preference, though some accounts of causation are committed to the conclusion that there is no true causality in microphysics.\footnote{David Papineau is a recent example: `{[}The{]} affinity with statistical mechanical accounts of thermodynamic behaviour argues that causation is an essentially macroscopic phenomenon, depending on structures which emerge only with respect to coarse-grained aspects of physical systems' (Papineau 2025, 14).} Again, an advantage of our shift from `collider' to `Correlator' is that it frees us from taking a stand on these issues, terminological or not.

These points do not give us a mechanism for the operational correlations on which our main proposal relies. But they do suggest some guidelines for the search for such mechanisms. First, we should be cautious, as Bell was, about the use of causal language, making sure we understand what we mean by it. Second, we should be \emph{more cautious} than Bell was about relying on intuitive time-asymmetries, no matter how straightforward they may seem. Again, it is widely accepted that the source of observed time-asymmetries lies not in fundamental theory but in a distant low entropy boundary condition. Unless we wish to challenge this orthodoxy -- or to accept that QM is not a fundamental theory -- we should take care to respect it, in discussing the ontological foundations of quantum phenomena.

\section{Constrained Correlators}

Our discussion above has skirted around a factor that makes a big difference to the \emph{character} of the correlations produced by holding fixed a Correlator -- that is, to the character of the Bell inequality-violating correlations themselves, according to the present proposal. Among other things, it may make a difference to the appropriateness of speaking about \emph{causality} across Bell experiments -- and hence, seemingly, take us back to where we began. In this final section we explain this factor, using toy models simpler than the actual Bell cases.

\subsection{Correlators via toy models}

To begin, let A and B be binary variables, and let D be a four-valued variable, correlated with A and B by the following rules: (i) each of the four possible joint values of AB is \textbf{incompatible} with the corresponding value of D; and (ii) the distributions of A, B and D are otherwise completely random. Figure 9 lists the possible values of these variables, and makes a visual allusion to the fact that D depends on A and B.

\begin{figure}[tbp]
\centering
\begin{tikzpicture}
    % nodes set with absolute coordinates
    \node[circle,draw] (A) at (0,0) {A};
    \node[circle,draw] (D) at (2,0) {D};
    \node[circle,draw] (B) at (4,0) {B};
\coordinate (coll) at (3,3); 
    % Directed edges
  \path[-,color=gray,rounded corners,line width=2pt] (A) edge (D);
   \path[-,color=gray,rounded corners,line width=2pt] (B) edge (D);

   \coordinate[label=right:{$(0,1)$}] (Ay) at (-0.5,-0.8);
   \coordinate[label=right:{(0,1)}] (By) at (3.55,-0.8);
   \coordinate[label=right:{$(00,01,10,11)$}] (Dy) at (0.8,-0.8);

\end{tikzpicture}
\caption{Three correlated variables} \label{fig:9}
\end{figure}

\begin{figure}[tbp]
  \centering

\begin{tikzpicture}%[remember picture,overlay]
\filldraw[draw=black,fill=white] (1,1) rectangle (2,2);
\filldraw[draw=black,fill=gray] (2,1) rectangle (3,2);
\filldraw[draw=black,fill=gray] (2,0) rectangle (3,1);
\filldraw[draw=black,fill=gray] (1,0) rectangle (2,1);

\filldraw[draw=black,fill=gray] (4,1) rectangle (5,2);
\filldraw[draw=black,fill=white] (5,1) rectangle (6,2);
\filldraw[draw=black,fill=gray] (5,0) rectangle (6,1);
\filldraw[draw=black,fill=gray] (4,0) rectangle (5,1);

\filldraw[draw=black,fill=gray] (1,-1) rectangle (2,-2);
\filldraw[draw=black,fill=gray] (2,-1) rectangle (3,-2);
\filldraw[draw=black,fill=gray] (2,-2) rectangle (3,-3);
\filldraw[draw=black,fill=white] (1,-2) rectangle (2,-3);

\filldraw[draw=black,fill=gray] (4,-1) rectangle (5,-2);
\filldraw[draw=black,fill=gray] (5,-1) rectangle (6,-2);
\filldraw[draw=black,fill=white] (5,-2) rectangle (6,-3);
\filldraw[draw=black,fill=gray] (4,-2) rectangle (5,-3);

\coordinate[label=right:{$D=00$}] (1) at (1.35,-0.3);
\coordinate[label=right:{$D=10$}] (2) at (4.35,-0.3);
\coordinate[label=right:{$D=01$}] (3) at (1.4,-3.3);
\coordinate[label=right:{$D=11$}] (4) at (4.4,-3.3);
\end{tikzpicture}
 
  \caption{D as a Correlator}
  \label{fig:10}
\end{figure}

These rules imply that for each value of D, the joint distributions of A and B are as depicted in Figure 10. It follows immediately that D is a Correlator with respect to A and B. Holding fixed D induces a correlation between A and B, as we saw above (\S2.1).

So far, we have not specified that A, B and D are physical variables with spacetime locations. If we now make that stipulation, we can consider various arrangements of these variables with respect to one another in time. In the following diagrams, we consider time to run in the vertical direction.

Let's first consider the case in which A and B are both \emph{earlier} than D (Figure 11). In this case, it is natural to suppose we can take control of the values of A and B, at least in principle. Representing A and B as intervention variables makes D into a collider, in the obvious way -- a common effect of independent causes A and B.

\begin{figure}[tbp]
  \centering

  \begin{tikzpicture}
    % nodes set with absolute coordinates
    \node[circle,draw] (A) at (0,0) {A};
    \node[circle,draw] (D) at (1.5,2) {D};
    \node[circle,draw] (B) at (3,0) {B};
%\coordinate (coll) at (3,3); 
    % Directed edges
  \path[-,color=gray,rounded corners,line width=2pt] (A) edge (D);
   \path[-,color=gray,rounded corners,line width=2pt] (B) edge (D);

\end{tikzpicture}

  \caption{D as a collider}
  \label{fig:11}
\end{figure}

Notice that if A and B are under the control of different experimenters, Alice and Bob, neither has any control over the choice of the other. They both make a difference to the value of D, but that's where their influence stops, so far as this simple model is concerned. These points may seem too trivial to mention, but it will be helpful to consider a strange variant of the case, where they can no longer be taken for granted.

\begin{figure}[bp]
  \centering

\begin{tikzpicture}
    % nodes set with absolute coordinates
    \node[circle,draw] (A) at (0,0) {A};
    \node[circle,draw, line width=2pt] (D) at (1.5,2) {D};
    \node[circle,draw] (B) at (3,0) {B};
%\coordinate (coll) at (3,3); 
    % Directed edges
  \path[-,color=gray,rounded corners,line width=2pt] (A) edge (D);
   \path[-,color=gray,rounded corners,line width=2pt] (B) edge (D);

   \coordinate[label=right:{$\mathbf{D=00}$}] (1) at (1.95,2);

\end{tikzpicture}

  \caption{D as a \emph{constrained} collider}
  \label{fig:12}
\end{figure}

\subsection{Adding `constraint'}

In Figure 12 we are supposing that there is some external boundary condition that \emph{fixes} or \emph{constrains} D to take a particular value -- as shown, the value 00. This may seem an absurd suggestion. Its main relevance will be in the time-reversed case, discussed below, in which D is earlier than A and B. There, the corresponding constraint turns out to be so ordinary that it easily escapes attention altogether.

Even in the present case (Figure 12), the idea that a future state of affairs should be fixed in this way isn't entirely unfamiliar. In fiction, even children's fiction, there are cases in which Fate fixes the future in some way, thereby giving characters some new ability to influence the present (e.g., Vernon 2015). In philosophy, Price and Weslake (2010) discuss an example called Death in Damascus. In this case, familiar in the relevant literature, an unfortunate hero is destined to meet Death tomorrow at noon. He has a choice where to do so, the traditional options being Aleppo and Damascus. Price and Weslake point out that the fact that the meeting is predestined gives the hero control over Death's present movements (at least if we assume that the hero really has a choice, and that Death is already on his way to the meeting point, whichever it is).

In the case in Figure 12, the effect is similar. If D is \emph{constrained} to take the value 00 then Alice can influence Bob's choice: if Alice chooses A=0, then Bob must choose B=1, because the combination 00 is disallowed when D=00. Of course, this sails into the seas of paradox. Why shouldn't Alice and Bob conspire to do something contradictory, by both choosing 0? This doesn't matter in the present example, which makes no claim to be realistic. The main point is to illustrate the effects of \emph{constraining} a collider, or more generally a Correlator. By restricting the options at D in this way, we induce a new dependency between A and B -- a correlation that begins to look a lot like causation, in that it can be used for control or signalling.

In case readers are getting impatient with this fantasy, we point out that there is at least one place in contemporary physics where something of this form has been seriously proposed. The Horowitz-Maldacena Hypothesis is a proposed solution to the black hole information problem. It rests on the proposal that there is an entanglement-swapping measurement at the future singularity, inside a black hole (Horowitz \& Maldacena 2004; Perry 2021). In our present terminology, this measurement is constrained to have a single result. This provides a path for in-falling information to couple to outgoing Hawking radiation, in what amounts to a constrained version of quantum teleportation.\footnote{Standard quantum teleportation would allow signalling and control, if we had access to entanglement swapping measurements thus constrained to yield a single outcome.}

As we said, the main relevance of Figure 12 is that it highlights something we are inclined to take for granted in the time-reversed case. Turning to that case, Figure 13 depicts the version of our toy model in which D is earlier in time than A and B.

\begin{figure}[tbp]
  \centering

\begin{tikzpicture}
    % nodes set with absolute coordinates
    \node[circle,draw] (A) at (0,0) {A};
    \node[circle,draw] (D) at (1.5,-2) {D};
    \node[circle,draw] (B) at (3,0) {B};
\coordinate (coll) at (3,3); 
    % Directed edges
  \path[-,color=gray,rounded corners,line width=2pt] (A) edge (D);
   \path[-,color=gray,rounded corners,line width=2pt] (B) edge (D);
\end{tikzpicture}
 
  \caption{D as an In-the-Past Correlator}
  \label{fig:13}
\end{figure}

The variable D is now an example of an In-the-Past Correlator, in the terminology of \S2.3. By controlling D we can guarantee a correlation between A and B (see Figure 14). We already have a name for this case: it is a \emph{constrained} Correlator. As promised, the notion that seemed fantastic in the In-the-Future case is perfectly ordinary here. \emph{Of course} we control initial conditions, in circumstances like this.

\begin{figure}[tbp]
  \centering

\begin{tikzpicture}
    % nodes set with absolute coordinates
    \node[circle,draw] (A) at (0,0) {A};
    \node[circle,draw, line width=2pt] (D) at (1.5,-2) {D};
    \node[circle,draw] (B) at (3,0) {B};
\coordinate (coll) at (3,3); 
    % Directed edges
  \path[-,color=gray,rounded corners,line width=2pt] (A) edge (D);
   \path[-,color=gray,rounded corners,line width=2pt] (B) edge (D);

   \coordinate[label=right:{$\mathbf{D=00}$}] (1) at (1.9,-2);

\end{tikzpicture}
 
  \caption{A \emph{constrained} In-the-Past Correlator}
  \label{fig:14}
\end{figure}

The most important lesson of this toy model is the one we learn by comparing Figures 13 and 14. In Figure 13 there is no need for a difference in A to be accompanied by a difference in B. Just as in Figure 11, such a difference can always be accommodated instead by a difference in D. But if D is constrained, as in Figure 14, this is no longer the case. Suppose A=1 and B=0, for example. Then a change in A would require a change in B, too, because D=00 rules out the case in which A=B=0.

We might say: `But that's true anyway, because D is in the past. A change in A can't produce a change in D -- that would be backward causation.' But this objection is itself backwards, in the present context. The reason we are inclined to dismiss backward causation may be that we take for granted that the past is constrained. One of the advantages of our toy model is that it enables us to relax that assumption, and think things through.

If we imagined a version of the In-the-Past case in which Alice and Bob could control the variables A and B, respectively, then all the lessons of the In-the-Future case would apply here too. If D is unconstrained then it is influenced both by A and by B -- D becomes a retrocausal collider, in effect. If D is constrained, then Alice now has influence over Bob's choice, and vice versa, and paradox looms, as before.

The point to stress is that in this simple toy model, the causation-like connection between A and B in Figure 14 arises from a combination of two things: (i) the dependence between D and the joint value of A and B; and (ii) the fact that D is constrained, in normal circumstances. If we allow ourselves to speak in relativistic terms, we could say that the \emph{spacelike} quasi-causal connection between A and B arises from two things: (i) the dependence of D on matters in its future light cone at A and B; and (ii) a constraint on D, of a kind that we have noted to be unremarkable in the In-the-Past case. In other words, we have something that behaves like a direct causal link between A and B. But it decomposes into relativity-friendly ingredients, in the way described.

Again, we might doubt the relevance of such an unrealistic model, but it is worth noting that it is unrealistic for a different reason than the previous case (Figure 12). There, it was the supposition that D was constrained \emph{in the future} that took us furthest from reality. In the present case, constrained correlators \emph{in the past} are unremarkable, as we noted in \S2.4. The unrealistic part of the story is that Alice and Bob might \emph{control} A and B, in the circumstances in which both variables remain correlated with D. In the classical cases imagined above, intervention by Alice and Bob would simply destroy the correlation.

\subsection{The non-classical character of Bell Correlators}

This brings us to an interesting contrast with the Bell case. The toy model of Figure 14 is simpler than the real Bell case, of course. Bell cases involve two variables at each side -- the setting and outcome, in the usual terminology -- of which Alice and Bob each control only one. But this difference in the number of variables isn't the most interesting contrast. The more interesting point is that -- unlike in any attempted classical implementation of the model of Figure 14 -- Alice and Bob can exercise this control \emph{without destroying the correlations on which the Correlating Fork depends.} Bell experiments are real-world In-the-Past Correlating Forks with (limited) experimental control at the two extremities, as well as a constrained Correlator at the vertex.

Unrealistic though it is, the toy model displays the kind of connections that might be expected to arise in such circumstances. Allowing some unreality on the QM side, we can imagine a Bell case in which Alice controls her outcome, as well as her setting.\footnote{This possibility arises in the de Broglie-Bohm view, if we have access to non-equilibrium ensembles; see Valentini 2002.} It would then be easy for her to signal. According to the proposal of this paper, the explanation of this imagined ability would be the same as in the toy model of Figure 14. In particular, it would depend on the fact that the initial state of the Bell experiment in question is \emph{constrained} -- held fixed to a particular value. Without this, control of both setting and outcome would not allow Alice to signal to Bob, because changes at \textbf{A} might make a difference at \textbf{C} (in the notation of Figure 6), with no difference at \textbf{B}.

All this maps over to the In-the-Future case in Figure 8. There, control of the outcome as well as the setting at \textbf{A} would not allow Alice to signal to \textbf{B}. Why not? Because changes at \textbf{A} could make a difference to the outcome at the entanglement-swapping measurement at \textbf{M}, with no need for a difference at \textbf{B}. Again, this changes if the outcome at \textbf{M} is constrained, as proposed by the Horowitz-Maldacena Hypothesis.

One last point about terminology. We have seen that in the presence of a constrained Correlator, Correlator Bias can give rise to a relationship between remote variables that behaves a lot like causation. Isn't this a threat to Bell's Principle of Local Causality (PLC)? As a reminder, Bell formulates that principle like this.

\begin{quote}
The direct causes (and effects) of events are near by, and even the indirect causes (and effects) are no further away than permitted by the velocity of light. (Bell 1990, 239)
\end{quote}
Yet we have just proposed a source of indirect spacelike causation, or something similar to it. Doesn't this take us back to the conclusion we promised to avoid, that Bell correlations are manifestations of some implausible kind of nonlocality? Doesn't it mean, in the tail-biting words of T S Eliot, that `the end of all our exploring/Will be to arrive where we started'?

However, what matters is not what we call our destination, but how we got here. Whatever we call these spacelike connections, two things are true. First, they depend on ingredients -- dependencies within lightcones, and ordinary control of initial states -- that do not themselves seem in tension with relativity. Second, they do not seem to have been on the radar in previous discussions of nonlocality, by Bell or anyone else.\footnote{We would be delighted to hear of exceptions.} The second point suggests that in deciding what to call the phenomenon, we needn't feel bound by previous usage. The first suggests that if we do decide to call it nonlocality, or even causality, we needn't conclude that it is something unwelcome. It may be true, in some sense, that this is where we started. But it is also true, in the same sense, that we now `know the place for the first time'.\footnote{This piece is greatly indebted to Ken Wharton. It builds on (Price \& Wharton 2025) and joint work over many years. I am also grateful to Carlo Rovelli, David Papineau, Richard Healey, Michael Hall, Arthur Fine, Simon Friedrich, Eric Cavalcanti, Howard Wiseman, Jason Grossman, Markus Müller, Gerard Ciepielewski, George Musser, Christian List, Lucas Boettcher, Joppe Widstam, Rob Spekkens, Wayne Myrvold, Naftali Weinberger, Zack Savitsky, and Travis Norsen; and to audiences at Griffith University, at Causal Worlds 2026, Grenoble, and at a workshop at MCMP, Munich in June 2026. I first heard the suggestion that preparation is preselection from Gerard Milburn. It took me some time to understand what he meant, and discussions in 2023 with Heinrich Päs and his group in Dortmund played a crucial role.}

\section*{References}
\begingroup
\small
\setlength{\parindent}{0pt}
\setlength{\parskip}{0.4em}
\everypar{\hangindent=1.5em\hangafter=1}

Arntzenius, F. 1990. Physics and common causes. \emph{Synthese} 82 (1), 77--96.

Bacciagaluppi, G. \& Hermens, R. 2021. Bell inequality violation and relativity of pre- and postselection. arXiv:2002.03935

Baez, John. 1996. A chat with Penrose. Blog post. \href{http://math.ucr.edu/home/baez/penrose.html}{math.ucr.edu/home/baez/ penrose.html}.

Bell, J. S. 1977. Free variables and local causality. \emph{Epistemological Letters} 15, 79--84. Reprinted in (Bell 2004), 100--104. Page references are to the latter version.

Bell, J. S. 1990. La nouvelle cuisine. In \emph{Between Science and Technology}, A. Sarlemijn and P. Kroes (eds.), 97--115. Amsterdam: Elsevier. Reprinted in (Bell 2004), 232--248. Page references are to the latter version.

Bell, J. S. 2004. \emph{Speakable and unspeakable in quantum mechanics}, 2nd edn. Cambridge: Cambridge University Press.

Carroll, Sean. 2017. Why Boltzmann Brains are Bad. \url{https://arxiv.org/abs/1702.00850}

Clauser, J. F., Horne, M. A., Shimony, A., and Holt, R. A. 1969. Proposed experiment to test local hidden-variable theories. \emph{Physical Review Letters}, 23: 880--884.

Dahlke, J. A. and Wiernik, B. M. 2019. Not Restricted to Selection Research: Accounting for Indirect Range Restriction in Organizational Research. \emph{Org Res Methods}, \emph{23}(4): 717-749.

Egg, M. 2013. Delayed-choice experiments and the metaphysics of entanglement. \emph{Foundations of Physics,} 43: 1124--1135.

Einstein, A., Podolsky, B. and Rosen, N. 1935. Can quantum-mechanical description of physical reality be considered complete? \emph{Physical Review} 47: 777--780.

Fankhauser, J. 2019. Taming the delayed choice quantum eraser. \emph{Quanta} 8, 44--56.

Friederich, Simon and Peter W. Evans. 2023. Retrocausality in Quantum Mechanics. \emph{The Stanford Encyclopedia of Philosophy} (Winter 2023 Edition).

Gaasbeek, B. 2010. Demystifying the delayed choice experiments. arXiv:1007.3977

Horowitz, G. and Maldacena, J., 2004. The black hole final state. JHEP 0402:008.

Keyes, Daniel. 1959. Flowers for Algernon. \emph{Fantasy and Science Fiction} 16(4): 5--30.

Mjelva, Jørn. 2024. Delayed-choice entanglement swapping experiments: no evidence for timelike entanglement. \emph{Stud.~Hist.~\& Phil.~Sc.} 105: 138--148.

Myrvold, Wayne, Genovese, Marco and Shimony, Abner. 2024. Bell's Theorem. \emph{The Stanford Encyclopedia of Philosophy (Summer 2024 Edition)}.

Papineau, D. 2025. Causal inference and the metaphysics of causation. \emph{Synthese} 206, 204.

Penrose, Roger. 1989. \emph{The Emperor's New Mind.} Oxford: Oxford University Press.

Penrose, Roger. 2004. \emph{The Road to Reality.} London: Jonathan Cape.

Perry, M. 2021. No future in black holes. arXiv:2106.03715

Price, Huw \& Weslake, Brad. 2010. The time-asymmetry of causation. In Beebee, Hitchcock and Menzies (eds), \emph{The Oxford Handbook of Causation} (OUP), 414--443.

Price, Huw \& Wharton, Ken. 2021. Entanglement swapping and action at a distance. \emph{Found.~of Phys.} 51: 105. \href{http://doi.org/10.1007/s10701-021-00511-3}{\nolinkurl{doi.org/10.1007/s10701-021-00511-3}}

Price, Huw \& Wharton, Ken. 2025. Taming entanglement. \href{https://arxiv.org/abs/2507.15128}{arxiv: 2507.15128}.

Shalm, L. K., Meyer-Scott, E., Christensen, B. G., Bierhorst, P., Wayne, M. A., Stevens, M. J., et al. Strong loophole-free test of local realism. \emph{Physical Review Letters} 115(25): 250402

%Trimmer, J. D. 1980. \emph{Proc American Philosophical Society,} 124: 3230--3338.

Valentini, A. 2002. Signal-locality in hidden-variables theories. \emph{Phys Lett A,} 297, 273--78.

Vernon, Ursula, 2015. \textit{Harriet the Invincible,} New York: Penguin Random House.

Weinberger, Naftali, 2026. Does the Past Hypothesis Explain the Causal Asymmetry? Preprint. \url{https://philsci-archive.pitt.edu/id/eprint/31188/}

Wharton, K. \& Argaman, N. 2020. Bell's Theorem and locally-mediated reformulations of quantum mechanics. \emph{Reviews of Modern Physics} 92, 21002. arXiv:1906.04313
\par\endgroup

\end{document}